\documentclass[prd, showpacs]{revtex4}
\usepackage{graphicx}
\usepackage{amssymb}
\usepackage{epsfig}
\usepackage{dcolumn}
\usepackage{bm}

\newcommand{\cc}{\cite}

\newcommand{\be}{\begin{equation}}
\newcommand{\ee}{\end{equation}}

\newcommand{\vecc}[1]{\mbox{\boldmath $#1$}}

\newcommand{\ba}{\begin{eqnarray}}
\newcommand{\ea}{\end{eqnarray}}
\newcommand{\bg}{\begin{gather}}
\newcommand{\foma}{\end{gather}}

\def\lagr{\hbox{$\cal L$}}

\def\pd{\partial}

\def\<{\langle}
\def\>{\rangle}

\def\a{\alpha}
\def\b{\beta}
\def\g{\gamma}  \def\G{\Gamma}
\def\d{\delta}  \def\D{\Delta}
\def\l{\lambda}   
\def\s{\sigma}
\def\r{\rho}

\def\m{\mu}
\def\n{\nu}
\def\t{\tau}

\def\({\left(}
\def\[{\left[}
\def\){\right)}
\def\]{\right]}

\def\pd{\partial}

\def\Tr{\hbox{Tr}}
\def\pa{{\cal P}}

\begin{document}

\title{\bf Feynman rules for effective Regge action}

\author{E.N. Antonov} \email{antonov@thd.pnpi.spb.ru}
\affiliation{\small Petersburg Nuclear Physics Institute,
RU-188350 PNPI, Gatchina, St.Petersburg, Russia}
\author{L.N. Lipatov}
\email{lipatov@thd.pnpi.spb.ru} \affiliation{\small Petersburg
Nuclear Physics Institute, RU-188350 PNPI, Gatchina,
St.Petersburg, Russia} \affiliation{Laboratoire de Physique Theorique et Hautes Energies,
Universite Pierre et Marie Curie, BP 126,
4, place Jussieu, F-75252 Paris, Cedex 05, France}
\author{E.A. Kuraev}\email{kuraev@thsun1.jinr.ru}
\affiliation{\small Joint Institute for Nuclear Research,
RU-141980 BLTP JINR, Dubna, Russia}
\author{I.O. Cherednikov} \email{igor.cherednikov@jinr.ru; igor.cherednikov@ca.infn.it}
\affiliation{\small
Joint Institute for Nuclear Research, RU-141980 BLTP JINR, Dubna,
Russia} \affiliation{Dipartimento di Fisica, Universit\`a di
Cagliari and INFN-Cagliari, C.P. 170, I-90142 Monserrato (CA),
Italy}

\date{\today}


\begin{abstract}
Starting from the gauge invariant effective action in the
quasi-multi-Regge kinematics (QMRK), we obtain the effective
reggeized gluon (R) -- particle (P) vertices of the following
types: $RPP$, $RRP$, $RRPP$, $RPPP$, $RRPPP$, and $RPPPP$,
where the {\it on}-mass-shell particles are gluons, or sets of
gluons with small invariant masses. The explicit expressions
satisfying the Bose-symmetry and gauge invariance conditions are
obtained. As a comment to the Feynman rules for derivation of the
amplitudes in terms of effective vertices we present a
``vocabulary'' for practitioners.
\end{abstract}

\maketitle

\section{Introduction}

The unitarization of the BFKL Pomeron as a composite state of two {\it reggeized gluons} is among the most
actual problems for investigations in QCD \cc{BFKL}. The BFKL evolution is applied to describe the
hadronic scattering amplitudes in the Regge regime $s \to \infty \
, \ s \gg -t $ and the high-energy behavior of total cross sections for semi-hard QCD processes ($Q^2\gg \Lambda ^2_{QCD}$).
In particular, the BFKL equation together with the DGLAP equation
\cc{DGLAP} are used to resum the logarithmic contributions to structure functions in the deep inelastic $ep$
scattering, and they are relevant to the gluon saturation at small Bjorken variable $x_{Bj}$---in particular, in
connection with the rapid growth of the function $F_2(x, Q^2)$, observed at HERA \cc{HERA}.

In the leading logarithmic approximation (LLA), the BFKL equation
predicts a power-like rise of cross sections with energy, which
obviously violates the Froissart constraint $\s_t < c \ln ^2s$
being a consequence of the $s$-channel  unitarity. One possible method  to overcome
this difficulty is to take into account the multiple Pomeron
exchanges in the eikonal approximation \cc{CW}. A more consistent
approach is based on the solution of the BKP equation which is a generalization of the BFKL equation 
for the case of composite states of several reggeized gluons \cc{BKP}. For large $N_c$, the BKP equation
turns out to be integrable \cc{int}. For the unitarization purpose, 
it is possible also to use an effective Lagrangian written for the multi-Regge kinematics
of gluons in intermediate states of the direct $s-$ and $u-$ channels
\cc{LPREV}. In this kinematics, the Feynman diagrams are factorized in the product of integrals 
over longitudinal and transverse subspaces. But the most general approach consists in reformulating QCD in
terms of a gauge-invariant effective field theory for the reggeized gluon interactions \cc{L95, LPHREP}.  Below we develop
this approach and apply it to the problem of calculating
production amplitudes in the quasi-multi-Regge kinematics (QMRK) for
the final particles.

The high energy scattering amplitudes can be presented in terms of
the Wilson $\pa$-exponentials \cc{NACH}. For a quasi-elastic kinematics, one
can derive a two dimensional sigma-model defined in the transverse subspace and constructed in terms of these
$\pa$-exponentials \cc{VERL}. However, this approach is incomplete and suffers from some
shortcomings (see \cc{L95, IA}). In the works by I. Balitsky
\cc{BAL}, the high-energy scattering was considered by means of
the amplitude factorization in the rapidity space within the shock-wave picture, and an hierarchy of
equations for scattering amplitudes was derived. The similar equations were obtained in the framework
of the JIMWLK approach \cc{JIMWLK}.
In these models, the high-energy
behavior in the large $N_c$ limit is described by a non-linear
evolution equation, now known as the Balitsky--Kovchegov (BK) equation
\cc{BAL, KOVCH}.
It has a simple interpretation in the framework of the dipole picture \cc{dip}. In turns, the dipole picture corresponds
to the M\"{o}bius representation for the Pomeron wave function, Ref. \cc{moeb}. The
similar, to some extent, methods based on the renormalization
group arguments were developed in Ref. \cc{KP}
(see also \cc{qq}).

On the other hand, the straightforward way to deal with the unitarization problem is to evaluate non-leading contributions
to the BFKL equation. The explicit calculation of the NLO logarithmic
corrections to inelastic amplitudes was performed in Ref. \cc{LF}
for one and two gluon production. Later on, the full result for the integral kernel of the BFKL equation in the next-to-leading
approximation was obtained \cc{FL_NLO, CC}. Although the corrections
are large, the modified equation can be used in the phenomenological applications (see, for example, Refs. \cc{bfklp, Altar, Ciaf}).

In the present study, we use the effective action approach proposed in Ref. \cc{L95}.
It appears to be convenient and physically well-motivated to perform the systematic self-consistent calculation
of non-leading contributions to the BFKL Pomeron intercept by means of an effective theory describing the interactions of
reggeized gluons with elementary particles (quarks and gluons). This approach describes
the processes with an arbitrary number of clusters of real and virtual
particles separated by large rapidity gaps, corresponding to QMRK for the produced particles.
For interactions local in the relative rapidities of particles the non-Abelian
gauge invariant effective action was derived \cc{L95}. This action contains
the fields of two sorts: the fields $A_{\pm}$, describing the reggeized gluons living  in the
$t$-channel, and the ordinary gluons and quarks produced and annihilated in the direct channels.

The motivation of this work is the necessity of developing of a proper practical tool for explicit calculations of the next-to-leading
corrections to the Feynman inelastic amplitudes of multi-peripheral
processes. The first thing one has to do before doing any calculations is to derive the Feynman rules corresponding to an
effective Lagrangian which is dealt with. In our case, such rules should prescribe the propagators of {\it reggeons} and ordinary
particles as well as the effective reggeon-particle vertices. A number of such effective vertices was already constructed
\cc{BFKL, L95, LF96, LPHREP} (see also \cc{VERT}). Note, that in Ref. \cc{LF} for the construction of production amplitudes the authors
used the string theory motivated approach \cc{STRING}.

Below we reproduce the known results and calculate new effective vertices within
the unified framework of the gauge invariant effective action
suggested in Ref. \cc{L95}. There is a hope that the formulation
of the Feynman rules derived from the effective action can provide a significant simplification for construction of 
the differential cross sections describing the creation of several
clusters of particles separated by rapidity gaps. We concentrate our
efforts on the presentation of the results in the form convenient for their use in numerical simulations and in
phenomenological applications.

The processes of such a kind can be investigated, {\it e.g.}, in experiments at RHIC and LHC. Another important branch of possible
applications---calculation of peripheral amplitudes with several
reggeized gluons in the scattering channel---is out of scope of the present paper and will not be considered here.

The structure of the paper is following: In the Section II, we fix notations, give an elementary ``vocabulary'' of terms and
kinematic constraints that are used in the work and remind the basic
relations for the effective action approach of Ref. \cc{L95}. In the Section III,
several Reggeon-Particle amplitudes are derived systematically within this approach and presented
in an explicit form. In the Section IV,  we discuss the calculation of the cross sections for the multi-jet production
process in the QMRK using the effective vertices derived in this
paper. In Conclusions, we summarize the results and propose possible applications of them in high energy phenomenology.

Some technical details are given in the Appendix.

\section{Effective action and recurrence relations for the reggeon vertices}

We consider the parton-parton collision at high energies $\sqrt{s}$ in
the center-of-mass system. The main contribution to the total
cross section stems from QMRK of final state particles (see Fig.1). In this regime, 
the final state particles compose several groups (clusters) with an arbitrary number of gluons or/and quarks with
a fixed mass $M_i$ ($i=1,2,...,n$) of each group. The clusters are produced in the multi-Regge kinematics with
respect to each other:
$$
P_A + P_B = Q_1 + Q_2 +... + Q_n \ , $$
$$
s = 2 P_A P_B = 4 E^2 \gg s_i= 2Q_iQ_{i+1} \gg |t_i| = |q_i^2| \ ,
\, i= 1, 2 , ... , n-1 \ ;
$$
\be
Q_i^2 = M_i^2 \ , \ Q_k = \sum _j p^{(k)}_j \ , \ k=1, 2, ..., n \ .
\label{qmrk}
\ee
Also, we introduce the light-cone vectors
\be 
n^+ = \frac{P_B}{E} \ \ , \ \
n^- = \frac{P_A}{E} \ \ , \ \ n^+ n^- = 2\ , \ (n^\pm)^2=0 
\ee 
and thus the light-cone projections of momenta and derivatives read,
respectively 
\be 
k^\pm = \(n^\pm\)_\m \cdot k^\m \ , \ \pd_{\pm} =
\(n^\pm\)_\m \cdot \pd^\m \ . 
\ee 
We imply also that the derivatives act on the particle $v(p)$ and reggeon $A(p)$ fields
in the momentum representation as follows:  \be \pd_\pm v(p) = - i
p_\pm v(p) \ , \ \frac{1}{\pd_\pm} v(p) =  \frac{i}{p_\pm} v(p) \ , \ 
\pd^2_\s A (q) = - q^2 A(q) \ ,\,\, q^2=-\vecc{q}^2_{\perp}. \ee
In the particular case, when there are only 3 clusters in the final
states (see Fig.1), we have two momentum transfers: 
\be 
q_1 = P_A - P_{A'} \ \ , \ \ q_2 = P_{B} - P_{B'} \ \ 
\ee 
with their Sudakov decompositions
\be 
q_1 = {q_1}_\perp + \frac{q_1^+}{2} \ n^- \  , \ q_2 =
{q_2}_\perp + \frac{q_2^-}{2} \ n^+ \ , \ q_1^- = q_2^+ = 0  \ .
\ee The Sudakov variables for produced particles are: \be p_i =
\frac{p_i^+}{2} \ n^- + \frac{p_i^-}{2} \ n^+ + {p_i}_\perp
\ . \ee
In the fragmentation regions, one has
\be
P_A+q_2 \to p_1 + p_2
+ ... + p_n  \ : \ P_A^+ = \sum_{i=1}^{n} p_i^+ \ , \ |q_2^+|
\ll p_i^+ \ , \ \label{cl1}
\ee
\be
\label{cl2}
 P_B+q_1 \to p_1 + p_2 + ... + p_n  \ : \ P_B^- = \sum_{i=1}^{n} p_i^- \ , \ |q_1^-| \ll p_i^- \ . \
\ee
In the central region one has
\be
\label{cl3}
q_1+q_2\to \sum_{i=1}^{n}p_i \ , \ q_1^+=
\sum_{i=1}^{n}p_i^+ \ ; \ q_2^-= \sum_{i=1}^{n}p_i^- \ .
\ee

To each reggeized gluon line (drawn in the crossing
channels $t_i$ in the Fig.1), one should attribute the sign $(+)$ (production)
to its end attached to initial parton with a large Sudakov component along $P_A$,
and the sign $(-)$ (annihilation) to the end attached to parton with a large Sudakov component along
$P_B$. Note, that wavy lines in Fig. 1 represent the reggeized gluons which are
{\it off}-mass-shell particles lying on the Regge trajectory. Initial,
final and produced particles (presented by the bold lines in the Fig. 1) 
are {\it on}-mass-shell particles considered to be massless. Let us emphasize that the presence of
reggeon lines within any production block is forbidden in  QMRK, Eq. (\ref{qmrk}).

To formulate the effective action one should introduce apart from the
usual fields $\psi (x),\bar \psi $ and $v_{\mu}(x)$ describing the quarks
and gluons, also the fields $A_{\pm}(x)$ describing the production
and annihilation of the reggeized gluons in the crossing channel. The quantities
$v_{\mu}(x), A_{\pm}(x)$ are implied to be anti-Hermitian matrices
belonging to the fundamental representation of the $SU(N)$ algebra:
\be
v_{\mu}=-iT^a v^a_{\mu}\,,\,\,A_{\pm}=-iT^a A^a_{\pm} \ \ , \ \ 
[T^a,T^b] = i f_{abc} T^c \ \ , \ \ \Tr \(T^a T^b\) = \frac{1}{2} \d^{ab} \ .
\ee
Here $T^a$ are Hermitian generators of the color group in the fundamental representation.

It is convenient to write down the effective action in the
following form \cc{L95}: 
\be 
S = \int\! d^4x \[\lagr_{QCD} + \lagr_{ind}\] \ \ , 
\ee 
where the standard Yang-Mills part consists of the quark-gluon and gluon-gluon interactions
\be
\lagr_{QCD} = i\bar \psi  \hat{D}\psi +
\frac{1}{2} \Tr \ G_{\mu\nu}^2 = \lagr_k + \lagr_{int} \ , \ \ D_{\mu} = \partial_{\mu} + g v_{\mu} \ , \ \ 
G_{\mu \nu} = \frac{1}{g} [D_{\mu},D_{\nu}] \ .
\ee
The explicit forms of kinetic and interacting parts of the QCD Lagrangian are:
$$ 
\lagr_k = i\bar{\psi}\hat{\partial}\psi-\frac{1}{4}(\partial_\mu v^a_\nu-\partial_\nu v^a_\mu)^2 \ ; $$ 
\be
\lagr_{int}=-g\bar{\psi}\hat{v}^a\psi+\frac{g}{2}f_{abc}(\partial_\mu v^a_\nu)v^b_\mu v^c_\nu-
\frac{g^2}{4}f_{abc}f_{ade}v^b_\mu v^c_\nu v^d_\mu v^e_\nu \ .
\ee
The induced part
\be
\lagr_{ind}=\lagr _{ind}^k+\lagr_{ind}^{GR} \ \ \ , \ \ \ 
\lagr _{ind}^k = -\partial _{\mu}A_+^a \cdot \pd_\m A_-^a
\ee
contains apart from the kinetic term $\lagr_{ind}^k$ also gluon-reggeon
couplings \cc{L95} 
$$ \lagr_{ind}^{GR} (v_\pm, A_\pm) =$$
$$
-\Tr \Bigg\{\frac{1}{g}\pd_{+} \ \[ \pa 
\exp \(-\frac{1}{2}\int ^{x^+}_{-\infty}v_+(x')dx'^+\) \] \cdot \pd^2_\s A_-(x) +
\frac{1}{g} \pd_- \ \[ \pa 
\exp \(-\frac{1}{2}\int ^{x^-}_{-\infty}v_-(x')dx'^-\) \] \cdot \pd^2_\s A_+(x)\Bigg\} =
$$
\be
\Tr \Bigg\{
\[v_+ - g v_+ \frac{1}{\pd_+}v_+ + g^2 v_+ \frac{1}{\pd_+}v_+
\frac{1}{\pd_+}v_+ - ... \] \pd^2_\s A_- +
\[v_- - g v_- \frac{1}{\pd_-}v_- + g^2 v_- \frac{1}{\pd_-}v_-
\frac{1}{\pd_-}v_- - ... \] \pd^2_\s A_+\Bigg\} \ , \label{eff_act} \ee
where the symbol 
$\pa$ orders the multiplication of matrices $v_{\pm}(x')$ in 
accordance with increasing of their arguments $x'^{\pm}$. Note, that in QMRK  
momenta of 
reggeons are transverse $-q_i^2 \approx {\vecc q_i}^2$ and the following kinematic relation
is implied  for the fields $$\pd_- A_+ = \pd_+ A_- = 0 \ , $$ which guaranties
the gauge-invariance of the action providing that the fields
$A_{\pm}$ are not changed under the local gauge transformations (see Refs. \cc{L95, LPHREP}).
Because $\lagr_{ind}^{GR}$ contains the linear term in $v_{\pm}$, the Euler-Lagrange
equations for the functional $S$ has a non-trivial classical solution $v_{\pm}=A_{\pm}+...$ 
even in the perturbation theory, which leads to a renormalization of the free action
for the fields $A_{\pm}$ \cc{L95, LPHREP}.

Going to the momentum space, one can define a sequence of the effective vertices
$\D^{\nu _0\nu _1 ...\nu _r +}_{a_0a_1...a_rc} (k^+_0,k^+_1,...,k^+_r)$
for the interaction of the field $A^c_+$ with ($r+1$) gluons with color indices
$a_0a_1...a_r$ and momenta $k_0,k_1,...,k_r$ and corresponding vertices
for the interaction of the field $A^c_-$ \cc{L95, LPHREP}.
Note, that for the usual and effective gluon vertices we imply all the momenta
to be {\it in-coming}. The Sudakov components $k^+_l$ satisfy the conservation law
$$
k^+_0+k^+_1+...+k^+_r=0 \ ,
$$
since the component $q^+$ of the reggeon momentum is negligibly small.

For example, for $r=0,1,2$ we have \cc{L95, LPHREP} $$
\D^{\nu_0\pm}_{a_0c}=i\<0|\lagr_{ind}^{GR}|v^{\nu_0}_{a_0}A^c_\pm(q)\>=iq^2(n^\pm)^{\nu_0}\delta^{a_0c}
\ ; $$ \be \D^{\nu _0\nu _1 +}_{a_0a_1c} (k^+_0,k^+_1)=i\<0|
\lagr_{ind}^{GR}|v^{\nu_0}_{a_0}(k_0) v^{\nu_1}_{a_1}(k_1) A^c_+(q)\>= -gq^2 \
f_{a_0a_1c} \, (n^+)^{\nu _1}\frac{1}{k_0^+}(n^+)^{\nu
_0}\,,\,\,\,k^+_0+k^+_1=0\,, \ee $$ \D^{\nu _0\nu _1\nu _2 +}_{a_0a_1a_2c}
(k^+_0,k^+_1,k^+_2)= {\vecc q}^{2}_{\perp}\, (n^+)^{\nu _0}(n^+)^{\nu
_1}(n^+)^{\nu _2} \left(\frac{f_{a_2a_0a}\,f_{a_1ac}}{k_1^+k_2^+}+
\frac{f_{a_2a_1a}\,f_{a_0ac}}{k_0^+k_2^+}\right)\,,\,\,\,k^+_0+k^+_1+k^+_2=0 \,
. $$ The last vertex is Bose-symmetric due to the Jacobi identity.

It is helpful to introduce the operators
$G_{a_0a_1...a_r} (k^+_0,k^+_1,...,k^+_r)$ using the definition
\be 
\D^{\nu _0\nu _1 ...\nu _r +}_{a_0a_1...a_rc} (k^+_0,k^+_1,...,k^+_r)=
-2q^2\,(n^+)^{\nu _0}(n^+)^{\nu _1}...(n^+)^{\nu _r}\,
\Tr \,\[ T^c G_{a_0a_1...a_r} (k^+_0,k^+_1,...,k^+_r)\] \, , \ r\geq 2 \ .
\ee
They can be constructed from the above effective action for arbitrary $r$ in
the generalized eikonal form
\be 
G_{a_0a_1...a_r} (k^+_0,k^+_1,...,k^+_r)=
\sum _{\{i_0,i_1,...,i_r\}}\frac{T^{a_{i_0}}
T^{a_{i_1}}T^{a_{i_2}}...T^{a_{i_r}}}{k_{i_0}^+(k^+_{i_0}+k^+_{i_1})
(k^+_{i_0}+k^+_{i_1}+k^+_{i_2})...(k^+_{i_0}+k^+_{i_1}+...+k^+_{i_{r-1}})} \ ,
\label{G}
\ee
where the sum is performed over all permutations of numbers $0,1,...,r$. Note,
that this expression is explicitly Bose symmetric.
Due to the gauge invariance of the effective action, the vertices $G$'s
satisfy the Ward identity in the form 
\be
k_r^+\,G_{a_0a_1...a_r} (k^+_0,k^+_1,...,k^+_r)=i\sum _{i=0}^{r-1} f_{a_ra_ia}
G_{a_0a_1...a_{i-1}aa_{i+1}...a_{r-1}} (k^+_0,k^+_1,...,k^+_{i-1},k^+_{i}+
k^+_r,k^+_{i+1},...,k^+_{r-1})\ ,
\ee
where we used the following relations for $i_t=r$
\be
k^+_r=\sum _{l=0}^tk^+_{i_l}-\sum _{l=0}^{t-1}k^+_{i_l}\,,\,\,
[T^{i_{r-1}},T^{r}]=if_{i_{r-1},i_r,a}T^a \ .
\ee
The above identity does not depend on the color representation of the generators $T^a$
in expression (\ref{G}) for $G$. It means, that we can use for the Wilson 
$\pa$-exponentials in eq. (\ref{eff_act}) an arbitrary representation providing that the matrices
$A_{\pm}$ are taken in the same representation and the corresponding trace normalization 
is chosen.

Thus, we obtain the following recurrent relations for the effective
vertices 
$\D^{\nu _0\nu _1 ...\nu _r +}_{a_0a_1...a_rc} (k^+_0,k^+_1,...,k^+_r)$  \cc{L95}: 
\be
\D^{\nu _0\nu _1 ...\nu _r +}_{a_0a_1...a_r} (k^+_0,k^+_1,...,k^+_r)=
\frac{i}{k_r^+}(n^+)^{\nu_r}\,\sum _{i=0}^{r-1} f_{a_ra_ia}
\D^{\nu _0\nu _1 ...\nu _{r-1} +}_{a_0a_1...a_{i-1}aa_{i+1}...a_{r-1}}
(k^+_0,k^+_1,...,k^+_{i-1},k^+_{i}+
k^+_r,k^+_{i+1},...,k^+_{r-1})\ .
\ee

In the next Section, we construct the Feynman rules for the reggeon vertices
in an agreement with the above recurrence relations.

\section{Vertices}

\subsection{Standard QCD Feynman rules}

The Yang-Mills interaction part of the effective action yields the
standard QCD Feynman rules (see Fig. 2a-c):
\be
{3g\hbox{-vertex}:} \ \ \   g f_{abc} \[(p_1 - p_2)_\l g_{\m\n} +
(p_2 - p_3)_\m g_{\n\l} + (p_3 - p_1)_\n g_{\l\m} \] \
 \equiv gf_{abc} \g_{\m\n\l}(p_1,p_2,p_3) \ , \label{YM1}
\ee
having the property
\be
\g_{\m\n\l} (p_1,p_2,p_3) = \g_{\l\m\n}(p_3,p_1,p_2) = - \g_{\n\m\l}(p_2,p_1,p_3)=... \ ,
\ee
$$
{4g\hbox{-vertex}:} \ \ \ i g^2
[f_{abl}f_{cdl}(g^{\m\s}g^{\n\l} - g^{\m\l}g^{\n\s}) +
f_{acl}f_{dbl}(g^{\m\n}g^{\l\s} -
g^{\m\s}g^{\l\n}) + $$
\be
+ f_{adl}f_{bcl}(g^{\m\l}g^{\s\n} - g^{\m\n}g^{\s\l}) \ \equiv
ig^2\g^{\m\n\l\s}_{abcd} \ , \label{YM2}
\ \g^{\m\n\l\s}_{abcd} = \g^{\n\m\l\s}_{bacd} = ...  \ ,
\ee
\be
\hbox{gluon propagator:} \ \ \ - i \d^{ab}\
\frac{g_{\m\n} }{k^2} \ , \label{YM3}
\ee
\be \hbox{quark-gluon vertex:} \ \ \ ig \bar u(p_1) \g^\m T^a u(p_2) \ ,
\label{qg}
\ee
\be
\hbox{quark
propagator:} \ \ \ i \frac{\hat p+m}{p^2 - m^2} \ . \label{qprop}
\ee
In addition, we specify the reggeized gluon propagator, Fig. 2d:
\be
-\frac{i}{2k^2}\d^{ab} \[ (n^+)^\m(n^-)^\n + (n^+)^\n(n^-)^\m \] \ .
\ee

\subsection{Effective $PPR$ vertices}

Knowledge of the induced and ordinary 3-vertices allows one to build
the effective 3-vertices which obey the Bose- and
gauge-symmetries. We use here the conservation laws in the form of Eqs. (\ref{cl1}, \ref{cl2}).

We distinguish four types of the $PPR$ vertices.
First group consists of the margin type vertices: $q_2 = -P_A + P_{A'} \ , \ q_1 = P_{B'} - P_B $.

\begin{itemize}

\item
``Left'' margin type (see Fig. 4a): $$ \g^{\n\n'+}_{\parallel abc}
\(P_A, a; P_{A'}, b; q_2,c\) = g f_{abc} \ \G^{\n\n'+}
\(P_A, q_2\) \ ,
$$ \be \G^{\n\n'+} \(P_A,q_2\)
= 2 P_A^+ g^{\n\n'} + \(n^+\)^\n \(-2P_A + P_{A'} \)^{\n'} +
\(n^+\)^{\n'} \(-2P_{A'} + P_{A}\)^{\n} - \frac{q_2^2}{P_A^+}
\(n^+\)^\n \(n^+\)^{\n'} \ . \ee One can check that the condition
of gauge invariance is explicitly satisfied:
\be
\G^{\n\n'+}(P_A,q_2) \cdot \(P_{A'}\)_{\n'} = P_A^\n P_{A}^+ -P_A^2(n^+)^\n
\, \ee \be \G^{\n\n'+}(P_A,q_2) \cdot \(P_{A}\)_{\n}  =
P_{A'}^{\n'} P_{A}^+ - P_{A'}^2 \(n^+\)^{\n'}  \ .
\ee
For {\it on}-mass-shell particle $A$: $P_A^2 = 0$ with the gauge condition
$\(e(P_{A}) \cdot P_{A} \) = 0 $ we have \be
\G^{\n\n'+} (P_A,q_2) \cdot \(P_{A'}\)_{\n'} \cdot e_\n
\(P_{A}\) = 0 \ . \ee

\item ``Right'' margin type (see Fig. 4b):
$$
\g^{\n\n'-}_{\parallel} \(P_B, a; P_{B'}, b; q_1,c \)
= -g f_{bac} \  \G^{\n\n'-} \(q_1,P_B\) \ ,
$$
\be
\G^{\n\n'-}  \(q_1,P_B\) = 2 P_B^-
g^{\n\n'} - \(n^-\)^{\n'} \(2P_{B'} - P_{B} \)^{\n} - \(n^-\)^{\n}
\(2P_{B} - P_{B'}\)^{\n'} - \frac{q_1^2}{P_B^-} \(n^-\)^\n
\(n^-\)^{\n'}  \ .
\ee
In this case, the gauge-invariance tests read
\be
\G^{\n\n'-} (q_1,P_B) \cdot \(P_{B'}\)_{\n'} =
P_B^\n P_{B}^- -P_B^2(n^-)^\n \ , \ee \be \G^{\n\n'-} (q_1,P_B) \cdot
\(P_{B}\)_{\n}  =  P_{B'}^{\n'} P_{B}^- -P_{B'}^2 \(n^-\)^{\n'} \ .
\ee

The second group includes the effective vertices of the {\it central}
type where $ \ k = q_1 + q_2 \  , \ q_{1,2}^2 \neq 0 \  $:

\item ``Left'' central type (see Fig. 4c):
$$ \g^{\n\n'+}_{\perp abc} \(q_1, a; k,
b; q_2,c \) = g f_{abc} \ \G^{\n\n'+} (q_1, q_2) \ , $$
\be \G^{\n\n'+} (q_1, q_2) = 2 q_1^+ g^{\n\n'} -
\(n^+\)^\n (q_1 - q_2)^{\n'} - \(n^+\)^{\n'} \(q_1 + 2 q_2\)^\n -
\frac{q_2^2}{q_1^+} \(n^+\)^\n \(n^+\)^{\n'}  \ . \label{LC}
\ee
The corresponding Ward identities read:
\be
\G^{\n\n'+} (q_1,q_2)
\cdot k_{\n'} = q_1^+ q_1^\n - q_1^2 \(n^+\)^\n  \ , \
\G^{\n\n'+} (q_1,q_2) \cdot \(q_1\)_\n =  q_1^+ k^{\n'} - k^2
\(n^+\)^{\n'} \ .
\ee

\item ``Right'' central type (see Fig. 4d):
$$ \g^{\n\n'-}_{\perp abc} \(q_1, a; k,
b; q_2,c \) = - g f_{abc} \ \G^{\n\n'-} (q_1, q_2)  \  ,$$
\be \G^{\n\n'-} (q_1, q_2) = 2 q_2^- g^{\n\n'} +
\(n^-\)^\n (q_1 - q_2)^{\n'} + \(n^-\)^{\n'} \(-q_2 - 2 q_1\)^\n -
\frac{q_1^2}{q_2^-} \(n^-\)^\n \(n^-\)^{\n'}  \ , \label{RC}
\ee
for which one has
\be \G^{\n\n'-} (q_1,q_2)\cdot k_{\n'} =  q_2^-
q_2^\n - \(n^-\)^\n q_2^2 \ , \ \G^{\n\n'-} (q_1,q_2) \cdot
\(q_2\)_\n = q_2^- k^{\n'} - k^2 \(n^-\)^{\n'} \ .
\ee

\end{itemize}

\subsection{Effective $PRR$ vertex}

Production of a single gluon with momentum $k_\m=\(q_1+q_2\)_\m $
and color index $b$ in the ``two reggeons collision'' in
color-momentum states, respectively, $\(q_1, \ a; \ k,\ b; \ q_2 \ c\)$,  is
described by the $PRR$ vertex (see Fig. 4e)
$$
\G^{-\m+}\(q_1, a; k, b ;q_2, c\) = g f_{abc} \ C^\mu(q_1,q_2) \ ,
$$
\be
f_{abc}C^\mu(q_1,q_2)=f_{abc}\g^{\n\mu\eta}(q_1,-k,q_2)(n^-)_\n(n^+)_\eta+
\Delta^{-\mu\eta}_{cba}(q_1,k,q_2)(n^+)_\eta-
\Delta^{\eta\mu+}_{abc}(q_1,k,q_2)(n^-)_\eta \ . \ee
As a result we have
\be
C^\mu(q_1,q_2)=2\[\(n^-\)^\mu\(q_1^++\frac{q_1^2}{q_2^-}\)-
\(n^+\)^\m\(q_2^-+\frac{q_2^2}{q_1^+}\)+\(q_2-q_1\)^\mu \ \] \ .
\ee The 4-vector $C^\mu$ obeys the gauge condition $k_\mu \cdot
C^\mu(q_1,q_2)  = 0$.

\subsection{Effective $RRPP$ vertex}

We consider first the case when pair of gluons in color-momenta
states $\( p_1\ , \ \n_1 \ , \ a_1; \ p_2 \ , \ \n_2 \ ,\ a_2 \)$
are created in collision of two reggeons with color-momenta states
$\( q_1 \ , \ c; \ q_2 \ , \ d\)$ with the momentum conservation
relation $q_1 + q_2 = p_1 + p_2$. It can be build in terms of the
effective 3-vertices given above (see Fig. 4) It has the form
$$
\frac{1}{i g^2}\G^{-\n_1\n_2+}_{ca_1a_2d}\(q_1;p_1,p_2;q_2\)= $$
$$ = \frac{T_1}{p_{12}^2}C^\eta\(q_1,q_2\)
\g^{\n_1\n_2\eta}\(-p_1,-p_2,p_{12}\)+
\frac{T_3}{\(p_2-q_2\)^2}\G^{\eta\n_1-}\(q_1,p_1-q_1\)\G^{\eta\n_2+}\(p_2-q_2,q_2\)-
$$ $$
\frac{T_2}{(p_1-q_2)^2}\G^{\eta\n_2-}\(q_1,p_2-q_1\)\G^{\eta\n_1+}\(p_1-q_2,q_2\)-
$$ $$ T_1\[\(n^-\)^{\n_1}\(n^+\)^{\n_2} -\(n^-\)^{\n_2}\(n^+\)^{\n_1}\]-
T_2\[2g^{\n_1\n_2}-\(n^-\)^{\n_1}\(n^+\)^{\n_2} \] -
T_3\[\(n^-\)^{\n_2}\(n^+\)^{\n_1}-2g^{\n_1\n_2}\] +
$$
\be
\D^{\rho\n_1\n_2+}_{ca_1a_2d}(q_1,p_1,p_2,q_2)(n^-)_\r
+ \D^{-\n_1\n_2\eta}_{ca_1a_2d}(q_1,p_1,p_2,q_2)(n^+)_\eta \ ,
\ee
where
\be
T_1=f_{a_1a_2r}f_{cdr} \ , \ T_2=f_{a_2cr}f_{a_1dr} \ , \ T_3=f_{ca_1r}f_{a_2dr} \ \ , \ \
T_1+T_2+T_3=0 \ , \label{jacobi1}
\ee
and the induced vertices, Fig. 3e:
\be
\D^{\rho\n_1\n_2+}_{ca_1a_2d}(q_1,p_1,p_2,q_2)(n^-)_\r = -
2q_2^2\(n^+\)^{\n_1}\(n^+\)^{\n_2}\(\frac{T_3}{p_2^+q_1^+} - \frac{T_2}{p_1^+q_1^+}\) \ ,
\ee
\be
\D^{-\n_1\n_2\eta}_{ca_1a_2d}(q_1,p_1,p_2,q_2)(n^+)_\eta =
-2q_1^2(n^-)^{\n_1}(n^-)^{\n_2}\(\frac{T_3}{p_1^-q_2^-} - \frac{T_2}{p_2^-q_2^-}\) \ .
\ee
One can verify the fulfillment of the gauge- and Bose-symmetries requirements:
\be
\G^{-\n_1\n_2+}_{ca_1a_2d}\(q_1;p_1,p_2;q_2\)p_{1{\n_1}} = 0 \ , \
\G^{-\n_1\n_2+}_{ca_1a_2d}\(q_1;p_1,p_2;q_2\) =
\G^{-\n_2\n_1+}_{ca_2a_1d}\(q_1;p_2,p_1;q_2\) \ .
\ee

\subsection{Effective $PPPR$ vertices}

Let us consider now the margin $PPPR$ vertices. The effective
$PPPR$ vertex can be constructed in terms of combination of
previously calculated effective vertices, and ordinary 3-vertices
(see Figs. 6, 7).

For the vertex describing the $A$-particle fragmentation region to
two gluons and a reggeon $P_A+R(q_2) \to g_1\(p_1\)+g_2\(p_2\)$
(see Fig. 6), one has
$$ \frac{1}{i
g^2}\G^{\nu_0\nu_1\n_2+}_{ca_1a_2d} \(P_A,p_1,p_2,q_2\) =
$$
$$ = \frac{T_1}{p_{12}^2}\g^{\s\n_1\n_2}(p_{12},-p_1,-p_2)
\G^{\n_0\s+}(P_A,q_2)+
\frac{T_3}{(p_2-q_2)^2}\g^{\n_0\n_1\s}(P_A,-p_1,p_1-P_A)
\G^{\s\n_2+}\(p_2- q_2,q_2\)+
$$
$$
\frac{T_2}{\(p_1-q_2\)^2}\g^{\n_0\n_2\s}(P_A,-p_2,p_2-P_A)
\G^{\s\n_1+}(p_1-q_2,q_2)+T_3\[g^{\n_1\n_2}
\(n^+\)^{\n_0}-(n^+)^{\n_1}g^{\n_0\n_2}\]+
$$
\be
T_2\[g^{\n_0\n_1}(n^+)^{\n_2}-g^{\n_1\n_2}(n^+)^{\n_0}\]+T_1\[(n^+)^{\n_1}
g^{\n_0\n_2}-\(n^+\)^{\n_2}g^{\n_0\n_1} \]+
\D^{\n_0\n_1\n_2+}_{ca_1a_2d}(p_A,p_1,p_2,q_2) \ ,
\ee
with the same definition for $T_i$'s as for the case $RRPP$ effective vertex, Eq. (\ref{jacobi1}),
and the conservation law $P_A + q_2 = p_1 + p_2$. When checking the gauge invariance
(convoluting with $\(p_1\)_{\n_1}$) we use the expression for the induced vertex (see Fig. 3d-left)
\be
\D^{\n_0\n_1\n_2+}_{ca_1a_2d}(p_A,p_1,p_2,q_2)=\frac{q_2^2}{P_A^+}
\[\frac{T_2}{p_1^+}-\frac{T_3}{p_2^+}\](n^+)^{\n_0}(n^+)^{\n_1}(n^+)^{\n_2} \ .
\ee
One can verify the fulfillment of the gauge condition:
\be
\G^{\n_0\n_1\n_2+}_{ca_1a_2d}\(P_A,p_1,p_2,q_2\)\cdot
\(p_1\)_{\n_1}=0 \ ,
\ee
as well as the Bose-symmetry condition:
\be
\G^{\n_0\n_1\n_2+}_{ca_1a_2d}\(P_A,p_1,p_2,q_2\)=
\G^{\nu_0\n_2\n_1+}_{ca_2a_1d}\(P_A,p_2,p_1,q_2\)=...  \ .
\ee

Similarly, we find for the ``right'' effective {\it margin} vertex, Fig. 7:
$$
\frac{1}{ig^{2}}\G^{-\n_1\n_2\n_0}_{ca_1a_2d}\(q_1,p_1,p_2,P_B\) =
$$
$$ -\frac{T_1}{p_{12}^2}\g^{\s\n_1\n_2}(p_{12},-p_1,-p_2)
\G^{\n_0\s-}(q_1,p_{12}-q_1)+\frac{T_3}{(p_1-q_1)^2}\g^{\s\n_2\n_0}
(p_2-P_B,-p_2,P_B) \G^{\s\n_1-}(q_1,p_1-q_1) - $$
$$  \frac{T_2}{(p_2-q_1)^2}\g^{\s\n_1\n_0}(p_1-P_B,-p_1,P_B)
\G^{\s\n_2-}(q_1,p_2-q_1)+T_1[(n^-)^{\n_2}g^{\n_0\n_1}-(n^-)^{\n_1}g^{\n_0\n_2}]+
$$
\be
T_3\[g^{\n_1\n_2}(n^-)^{\n_0}-(n^-)^{\n_2}g^{\n_0\n_1}\] +
T_2\[\(n^-\)^{\n_1}g^{\n_0\n_2}-\(n^-\)^{\n_0} g^{\n_1\n_2}\] +
\D^{-\n_1\n_2\n_0}_{ca_1a_2d}(q_1,p_1,p_2,P_B) \ ,
\ee
with the same notations for $T_i$'s as for ``left'' margin vertex and the $RRPP$ vertex,
Eq. (\ref{jacobi1}), conservation law $q_1 + P_B = p_1 + p_2$, and the induced vertex,
Fig. 3d-right:
\be
\D^{-\n_1\n_2\n_0}_{ca_1a_2d}(q_1,p_1,p_2,P_B)=\frac{q_1^2}{P_B^-}\[
\frac{T_2}{p_2^-}-\frac{T_3}{p_1^-}\](n^-)^{\n_0}(n^-)^{\n_1}(n^-)^{\n_2} \ .
\ee
Again, the gauge condition is fulfilled:
\be
\G^{-\n_1\n_2\n_0}_{da_1a_2c}\(q_1,p_1,p_2,P_B\) \cdot
\(p_1\)_{\n_1}=0 \ , \ee
as well as the Bose-symmetry.

\subsection{Effective $RRPPP$, and $RPPPP$ vertices}

\begin{itemize}

\item Consider first the $RRPPP$ vertex $V^{+\n_1\n_2\n_3-} (q_1,
p_1,p_2,p_3,q_2)$. It can be constructed by means of combinations
of the already known effective vertices given above, Eqs.
(\ref{YM1}, \ref{YM2}, \ref{YM3}, \ref{LC}, \ref{RC}).
The result is graphically presented at the Fig. 8. Each of these
diagrams presents, indeed, a set of diagrams related to each other
by any possible permutations of the produced particles.
Using the rules given in the Appendix, one can write down the analytical
expression summing up all these contributions:
$$
\frac{1}{g^3}\G_{d123e}^{-\n_1\n_2\n_3+} (q_1, p_1,p_2,p_3,q_2) =
\frac{f_{edr}}{k^2} C_\s(q_1,q_2) \cdot A_{a_1a_2a_3r}^{\n_1\n_2\n_3\s} + $$
$$
\[f_{mdr}f_{r13}f_{em2} \frac{1}{p_{13}^2(p_2-q_2)^2}
\G^{\s\eta-}(q_1,p_{13}-q_1)\G^{\s\n_2+}(p_2-q_2,q_2)\g^{\n_1\n_3\eta}(-p_1,-p_3,p_{13})+
2 \ \hbox{\tt{perm.}} \] + $$
$$
\[f_{md3}f_{r12}f_{emr}\frac{1}{p_{12}^2(p_3-q_1)^2}
\G^{\s\n_3-}(q_1,p_3-q_1)\G^{\s\eta+}(p_{12}-q_2,q_2)\g^{\n_1\n_2\eta}
(-p_1,-p_2,p_{12})+ 2\ \hbox{\tt{perm.} } \] +
$$
$$
\[f_{md1}f_{rm2}f_{er3}\frac{1}{(q_1-p_1)^2(p_3-q_2)^2}
\G^{\s\n_1-}(q_1,p_1-q_1)\G^{\eta\n_3+}(p_3-q_2,q_2)
\g^{\s\n_2\eta} (q_1-p_1,-p_2,q_2-p_3)+ 5 \ \hbox{\tt{perm.}}\] + $$
$$
\[f_{er3}\frac{1}{(p_3-q_2)^2}\G^{\s{\n_3}+}(p_3-q_2,q_2)\(\D + \g\)^{-\n_1\n_2\s}_{d12r}
(q_1, p_1, p_2, q_2-p_3 ) + 2 \ \hbox{\tt{perm.} } \] - $$
$$
\[f_{rd3}\frac{1} {(p_3-q_1)^2}\G^{\s\n_3-}(q_1,p_3-q_1)\(\D + \g \)^{\s\n_1\n_2+}_{r12e}
(q_1-p_3,p_1,p_2,q_2)+ 2 \ \hbox{\tt{perm.} } \] + $$
$$
\[f_{r23}\frac{1}{p_{23}^2}\g^{\n_2\n_3\s}(-p_2-p_3,p_{23})\(\D^{-\s\n_1\eta}_{dr1e}\cdot
(n^+)_\eta + \g^{-\s\n_1+}_{dr1e} +
\D^{\rho\s\n_1+}_{dr1e}\cdot(n^-)_\r \)(q_1,p_{23},p_1,q_2)+ 2 \ \hbox{\tt{perm.} } \] + $$
\be
\(\D^{-\n_1\n_2\n_3\rho}_{d123e}\cdot(n^+)_\rho+
\D^{\rho\n_1\n_2\n_3+}_{d123e}\cdot(n^-)_\rho\) (q_1,p_1,p_2,p_3,q_2) \ ,
\ee
where we use the shorthand notations
\be f_{123} \equiv f_{a_1a_2a_3} \ , \ p_{12}=p_1+p_2 \ , \ k=p_1+p_2+p_3 \ . \ee

For the induced terms of the rank-2 entering $RRPPP$ we have (see Fig. 3g):
\be
\D^{-\n_1\n_2\s}_{d12r}(q_1,p_1,p_2,q_2-p_3)=-\frac{q_1^2}{p_1^-}(n^-)^{\n_1}
(n^-)^{\n_2}(n^-)^{\s} \[\frac{f_{1rm}f_{2dm}}{p_2^-}+\frac{f_{rdm}f_{12m}}{p_{12}^-} \] \ ,
\ee
\be
\D^{-\s\n_1\eta}_{dr1e}(q_1,p_{23},p_1,q_2)=-\frac{q_1^2}{p_1^-}(n^-)^{\n_1}
(n^-)^{\eta}(n^-)^{\s}\[\frac{f_{1em}f_{drm}}{p_{23}^-}+\frac{f_{edm}f_{1rm}}{q_2^-}\] \ ,
\ee
\be
\D^{\s\n_1\n_2+}_{r12e}(q_1-p_3,p_1,p_2,q_2)=\frac{q_2^2}{p_1^+}(n^+)^{\n_1}
(n^+)^{\n_2}(n^+)^{\s} \[\frac{f_{12m}f_{erm}}{p_{12}^+}+\frac{f_{2em}f_{1rm}}{p_2^+}\] \ ,
\ee
\be
\D^{\rho\s\n_1+}_{dr1e}(q_1,p_{23},p_1,q_2)=\frac{q_2^2}{p_1^+}(n^+)^{\n_1}
(n^+)^{\rho}(n^-)^{\s}\[\frac{f_{d1m}f_{erm}}{p_{23}^+}+\frac{f_{edm}f_{1rm}}{q_1^+}\] \ ,
\ee
with conservation law
$$
q_1^+ = p_1^+ + p_2^+ + p_3^+ \ , \ q_2^-=p_1^-+p_2^-+p_3^- \ .
$$

\item For the ``left'' {\it margin} effective vertex of the rank-3 we have:
$$
\frac{1}{g^3}\G_{d123e}^{\rho\n_1\n_2\n_3+} (P_A,
p_1,p_2,p_3,q_2) = \frac{f_{edr}}{k^2} \G^{\rho\s+}(P_A,q_2) \cdot
A_{r123}^{\s\n_1\n_2\n_3} -
\[f_{md1}f_{rm2}f_{er3}\frac{1}{(P_A-p_1)^2(p_3-q_2)^2} \cdot \right. $$
$$ \left.
\G^{\eta\n_3+}(p_3-q_2,q_2)\g^{\s\n_2\eta}(P_A-p_1,-p_2,q_2-p_3)
\g^{\rho\n_1\s}(P_A,-p_1,p_1-P_A) + 5 \ \hbox{\tt{perm.}} \] -
$$
$$
\[f_{rdm}f_{m12}f_{er3} \frac{1}{p_{12}^2(p_3-q_2)^2}
\G^{\eta\n_3+}(p_3-q_2,q_2)\g^{\rho\s\eta}(P_A,-p_{12},q_2-p_3)\g^{\s\n_1\n_2}
(p_{12},-p_1,-p_2)+ 2 \ \hbox{\tt{perm.}} \] + $$
$$
\[f_{md1}\frac{1}{(p_1-P_A)^2}
\g^{\rho\n_1\s}(P_A,-p_1,p_1-P_A)\[-\frac{1}{p_{23}^2}\G^{\s\eta+}(p_{23}-q_2,q_2)
\g^{\n_2\n_3\eta}(-p_2,-p_3,p_{23})f_{emr}f_{r23} + \right. \right. $$
$$
\left. \left. + \(\D + \g\)^{\s\n_2\n_3+}_{m23e}(P_A-p_1,p_2,p_3,q_2)\] + 2 \ \hbox{\tt{perm.}} \] +
\[f_{er1}\frac{1}{(p_1-q_2)^2}\G^{\s\n_1+}(p_1-q_2,q_2) \gamma^{\rho\n_2\n_3\s}_{d23r} +
2 \hbox{\tt{perm.}} \] + $$
\be
\[\frac{f_{r23}}{p_{23}^2}\g^{\n_2\n_3\eta}(-p_2,-p_3,p_{23})
\(\D + \g\)^{\rho\eta\n_1+}_{dr1e}(P_A,p_{23},p_1,q_2)+ 2 \ \hbox{\tt{perm.}} \] -
\D^{\rho\n_1\n_2\n_3+}_{d123e}(P_A,p_1,p_2,p_3,q_2) \ ,
\ee
where the ``left'' {\it margin} induced vertex, Fig. 3d-left, reads
\be
\D^{\s\n_2\n_3+}_{m23e}(P_A-p_1,p_2,p_3,q_2)=\frac{q_2^2}{p_{23}^+}(n^+)^{\s}
(n^+)^{\n_2}(n^+)^{\n_3}\(\frac{f_{3er}f_{emr}}{p_3^+}-\frac{f_{e2r}f_{3mr}}{p_2^+}\) \ ,
\ee
and
\be
\D^{\rho\eta\n_1+}_{dr1e}(P_A,p_{23},p_1,q_2)=\frac{q_2^2}{p_1^+}(n^+)^{\rho}
(n^+)^{\n_1}(n^+)^{\eta}\(-\frac{f_{erm}f_{1dm}}{p_{23}^+}-\frac{f_{r1m}f_{edm}}{P_A^+} \) \ ,
\ee
and the conservation law $P_A^+=p_1^++p_2^++p_3^+$ is implied.

\item The ``right'' margin effective vertex is
$$
\frac{1}{g^3}\G_{e123d}^{-\n_1\n_2\n_3\rho} (q_1,
p_1,p_2,p_3,P_B) =-\frac{f_{der}}{k^2} \G^{\rho\s-}(q_1,P_B) \cdot
A_{r123}^{\s\n_1\n_2\n_3} +
\[f_{me1}f_{rm2}f_{dr3}\frac{1}{(P_B-p_3)^2(p_1-q_1)^2} \cdot \right. $$
$$
\left.
\G^{\s\n_1-}(q_1,p_1-q_1)\g^{\eta\s\n_2}(P_B-p_3,q_1-p_1,-p_2)
\g^{\rho\eta\n_3}(P_B,p_3-P_B,-p_3) + 5 \ \hbox{\tt{perm.}} \] + $$
$$
\[f_{drm}f_{m12}f_{re3} \frac{1}{p_{12}^2(p_3-q_1)^2}
\G^{\s\n_3-}(q_1,p_3-q_1)\g^{\rho\s\eta}(P_B,p_{12}-P_B,-p_{12})
\g^{\eta\n_1\n_2} (p_{12},-p_1,-p_2)+2 \ \hbox{\tt{perm.}} \] + $$
$$
\[f_{dr1}\frac{1}{(p_1-P_B)^2}
\g^{\rho\eta\n_1}(P_B,p_1-P_B,-p_1)\[\frac{f_{rem}f_{m23}}{p_{23}^2}\G^{\eta\s-}(q_1,p_{23}-q_1)
\g^{\s\n_2\n_3}(p_{23},-p_2,-p_3)+ \right. \right.$$
$$ \left. \left.
\(\D + \g\)^{-\n_2\n_3\eta}_{e23r}(q_1,p_2,p_3,P_B-p_1)\]+ 2 \ \hbox{\tt{perm.}} \] - $$
$$
\[f_{me1}\frac{1}{(p_1-q_1)^2}\G^{\s\n_1-}(q_1,p_1-q_1)\g^{\s\n_2\n_3\rho}_{m23d} +
2 \ \hbox{\tt{perm.} } \] + $$
\be
\[\frac{f_{m23}}{p_{23}^2}\g^{\s\n_2\n_3}(p_{23},-p_2,-p_3)
\(\D + \g \)^{-\s\n_1\rho}_{em1d}(q_1,p_{23},p_1,P_B)+ 2 \ \hbox{\tt{perm.}} \] -
\D^{-\n_1\n_2\n_3\rho}_{e123d}(q_1,p_1,p_2,p_3,P_B) \ ,
\ee
with the induced vertex, Fig. 3d-right:
\be
\D^{-\n_2\n_3\eta}_{e23r}(q_1,p_2,p_3,P_B-p_1)=\frac{q_1^2}{p_{23}^-}(n^-)^{\n_2}
(n^-)^{\n_3}(n^-)^{\eta}\[\frac{f_{3em}f_{2rm}}{p_3^-}-\frac{f_{e2m}f_{3rm}}{p_2^-}\] \ ,
\ee
and
\be
\D^{-\s\n_1\rho}_{em1d}(q_1,p_{23},p_1,P_B)=\frac{q_1^2}{p_1^-}(n^-)^{\n_1}
(n^-)^{\rho}(n^-)^{\s}\[\frac{f_{edr}f_{1mr}}{P_B^-}-\frac{f_{d1r}f_{emr}}{p_{23}^-} \] \ ,
\ee
and the conservation law $P_B^-=p_1^-+p_2^-+p_3^- \ . $

\end{itemize}

Here by term ``$\tt{perm.}$'' we imply all the terms obtained from the explicitly written
ones by simultaneous permutations of the momenta, color and Lorentz
indices. Like in the previous cases, the gauge invariance and Bose-symmetry properties are
fulfilled explicitly. Some useful relations which can be used are presented in the Appendix.

The ``left'' induced vertex of the rank-3 has a form
\be
\D_{d123e}^{\r\n_1\n_2\n_3+}(q_1,p_1,p_2,p_3,q_2) = q_2^2 \(n^+\)^{\r}
\(n^+\)^{\n_1}\(n^+\)^{\n_2}\(n^+\)^{\n_3}
\frac{D^+(q_1,d;p_1,a_1;p_2,a_2;p_3,a_3;q_2,e)}{p_1^+} \ ,
\ee
where
\be
\label{ind3}
D^+(q_1,d;p_1,a_1;p_2,a_2;p_3,a_3;q_2,e)=\frac{1}{p_{23}^+}
\[\frac{\alpha_1}{p_2^+}-\frac{\alpha_2}{p_3^+}\]+
\frac{1}{p_{12}^+}\[\frac{\alpha_3}{q_1^+}+\frac{\alpha_4}{p_3^+}\] +
\frac{1}{p_{13}^+}\[\frac{\alpha_5}{q_1^+}+\frac{\alpha_6}{p_2^+}\] \ ,
\ee
and
$$
\alpha_1=f_{md1}f_{e2r}f_{3mr} \ ; \quad \alpha_2=f_{md1}f_{3er}f_{2mr} \ ;
\quad \alpha_3=f_{m12}f_{der}f_{3mr} \ ; $$
\be
\alpha_4=f_{m12}f_{e3r}f_{dmr} \ ; \quad \alpha_1=f_{m13}f_{der}f_{2mr} \ ;
\quad \alpha_6=f_{m13}f_{e2r}f_{dmr} \ .
\ee
For the ``right'' induced vertex of the rank-3 (Fig. 3g-right) we have
\be
\D_{da_1a_2a_3e}^{-\n_1\n_2\n_3\r}(q_1,p_1,p_2,p_3,q_2) =
q_1^2\(n^-\)^{\n_1}\(n^-\)^{\n_2}\(n^-\)^{\n_3}\(n^-\)^{\r}
\frac{D^-(q_1,d;p_1,a_1;p_2,a_2;p_3,a_3;q_2,e)}{p_1^-} \ ,
\ee
where
\be
D^-(q_1,d;p_1,a_1;p_2,a_2;p_3,a_3;q_2,e)=\frac{1}{p_{23}^-}
\(\frac{\gamma_1}{p_3^-}-\frac{\gamma_2}{p_2^-}\)-
\frac{1}{p_{12}^-}\(\frac{\gamma_3}{q_2^-}+\frac{\gamma_4}{p_3^-} \) -
\frac{1}{p_{13}^-}\(\frac{\gamma_5}{q_2^-}+\frac{\gamma_6}{p_2^-} \) \ ,
\ee
and
$$
\g_1=f_{m1e}f_{3dr}f_{2mr} \ ; \ \quad \gamma_2=f_{m1e}f_{d2r}f_{3mr} \ ; \
\quad \gamma_3=f_{m12}f_{der}f_{3mr} \ ; $$
\be
\g_4=f_{m12}f_{3dr}f_{emr}\ ; \ \quad \gamma_5=f_{m13}f_{der}f_{2mr} \ ; \
\quad \gamma_6=f_{m13}f_{2dr}f_{emr} \ .
\ee

The Bose-symmetry can be proven similarly to the case of the ``right'' induced vertex
of the rank-4, if one takes into account the relation
\be
\D_{ea_1a_2a_3d}^{-\n_1\n_2\n_3\r}(q_1,p_1,p_2,p_3,q_2)=\D_{da_1a_2a_3e}^{\r\n_1\n_2\n_3+}
(q_1,p_1,p_2,p_3,q_2)|_{p_i^+\to p_i^- \ ; \ d\to e \ ; \ e\to d \ ; \ q_1^+\to q_2^-} \ .
\ee
Besides this, let us note that the {\it margin} induced vertices of the rank-3 are connected with
the {\it central} ones:
\be
\D_{da_1a_2a_3e}^{-\n_1\n_2\n_3\r}(q_1,p_1,p_2,p_3,P_B)=
\D_{ea_1a_2a_3d}^{-\n_1\n_2\n_3\r}(q_1,p_1,p_2,p_3,q_2)|_{q_2\to P_B\ ; \ d\to e\ ;\ e\to d} \ ,
\ee
\be
\D_{ea_1a_2a_3d}^{\rho\n_1\n_2\n_3+}(P_A,p_1,p_2,p_3,q_2)=
\D_{ea_1a_2a_3d}^{\rho\n_1\n_2\n_3+}(q_1,p_1,p_2,p_3,q_2)|_{q_1\to P_A} \ .
\ee

The effective 4-gluon vertex $A$ entering to RRPPP and RPPPP vertices reads
$$
A_{a_1a_2a_3r}^{\n_1\n_2\n_3\s} = \tau_1
\[g_{\n_1\s}g_{\n_2\n_3} - g_{\n_1\n_3}g_{\n_2\s} \] +
\tau_2\[g_{\n_1\n_2}g_{\n_3\s} - g_{\n_1\s}g_{\n_2\n_3}
\] +
$$
$$
\tau_3 \[g_{\n_1\n_3}g_{\n_2\s} - g_{\n_1\n_2}g_{\n_3\s} \]
+ \frac{\tau_3}{p_{23}^2} \g_{\s\n_1\r} (k,
-p_1,-p_2-p_3) \g_{\r\n_2\n_3} (p_2+p_3,-p_2,-p_3) +
$$
$$
\frac{\tau_2}{p_{13}^2} \g_{\s\r\n_2}(k,
-p_1-p_3,-p_2) \g_{\r\n_1\n_3} (p_1+p_3,-p_1,-p_3) -
$$
\be -
\frac{\tau_1}{p_{12}^2} \g_{\s\r\n_3}(k,
-p_1-p_2,-p_3) \g_{\r\n_1\n_2}(p_1+p_2, -p_1,-p_2) \ ,
\ee
where
$$
\t_1=f_{12m}f_{3rm} \ , \ \t_2=f_{31m}f_{2rm} \ , \ \t_3=f_{23m}f_{1rm} \ , \ \t_1+\tau_2+\tau_3=0 \ . $$

\section{Cross sections for jet production in QMRK}

The typical process of the $n$-jet production in the
quasi-multi-Regge kinematics is presented graphically at Fig. 1.
Within this regime, one can use the Sudakov decomposition
\be
q_i = \a_i P_B + \b_i P_A + {\vecc q_i}_\perp \ \  , \ \ {\vecc
q_i}_\perp \cdot P_A = {\vecc q_i}_\perp \cdot P_B =0 \ , \ d^4
q_i = \frac{s}{2} d\a_i d\b_i d^2 {\vecc q_i} \ , \label{sud}
\ee
and the invariant masses of jets:
$$ \(P_A - q_1\)^2 = M_1^2 \approx - s \a_1 \ ,
$$
$$ \(q_1 - q_2\)^2  = M_2^2 \approx - s \a_2\b_1 \ , $$
$$ \ \ \ ... \ \ \ $$
$$ \(q_{n-2} - q_{n-1}\)^2  = M_{n-1}^2 \approx - s \a_{n-1}\b_{n-2} \ , $$
\be \(q_{n-1}+ P_B\)^2  = M_n^2 \approx  s \b_{n-1} \ ,
\label{mass} \ee by virtue of the ordering valid in the QMRK: \be
\a_1 \ll \a_2 \ll ... \ll \a_n \ , \ \b_{n-1} \ll \b_{n-2} \ll ...
\ll \b_1 \ , \ M_i^2 \sim M^2 \ll s \ , \ee \be s_1 = (P_A -
q_2)^2 \sim s_2 = (q_1 -q_3)^2 \sim ... \sim s_{n-1} =
(q_{n-2} + P_B)^2 \gg M^2\ . \ee Here $M_i$ is the effective mass of the
$i-$th jet.
These quantities are related as $\prod_1^{n-1}s_i=s\prod_2^{n-1}M_i^2$.

The differential cross section of the $n$-jet production is given by \be d\s^{2 \to n_{jet}} = \frac{1}{8s}
|{\cal M}^{2\to n }|^2 d\G_n^{(jet)} \ . \label{csn} \ee
The phase volume for the $n$ jets containing $n_i \ , \ i = 1, ..., n$
produced particles in each jet can be written as \be d\G^{(jet)}_n
= \frac{(2\pi)^4}{s \cdot 2^{n-1}} \prod_{i=1}^{n-1} d^2{\vecc
q_i} \prod_{i=1}^{n}d\phi_i \ \int_{\frac{M^2}{s}}^{1} \!
\frac{d\b_1}{\b_1} \int_{\b_1}^{1} \! \frac{d\b_2}{\b_2} \cdot
\cdot \cdot \int_{\b_{n-3}}^{1} \! \frac{d\b_{n-2}}{\b_{n-2}} =
\frac{(2\pi)^4}{s \cdot 2^{n-1}} \prod_{i=1}^{n-1} d^2{\vecc q_i}
\prod_{i=1}^{n} d\phi_i \cdot \frac{(\ln
\frac{s}{M^2})^{n-2}}{(n-2)!} \ , \
\ee
where
$$
d\phi_1 = dM_1^2 \ d\G_1(2\pi)^4 \d^{(4)} \(P_A - q_1 - \sum_{j=1}^{n_1}p_j^{(1)}\) \ ,
$$
$$ d\phi_i = dM_i^2 \ d\G_i(2\pi)^4 \d^{(4)} \(q_{i-1} - q_i - \sum_{j=1}^{n_2}p_j^{(i)}\) \ ,
i=2,...,n-1 \ ,
$$
\be
d\phi_n = dM_n^2 \,
d\G_n(2\pi)^4 \d^{(4)} \(q_{n-1} + P_B - \sum_{j=1}^{n_n}p_j^{(n)}\) \ ,
\ee and \be d\G_i = \prod_{j=1}^{(n_i)} \frac{d^3 p_j^{(i)}}{2
\varepsilon_j (2\pi)^3} \ .
\ee
Therefore, taking into account
that the amplitude for $2 \to n$ production is given by
\be {\cal
M}^{2 \to n} = 2^{-n+1} s\cdot g^{n} \ V_A V_B
\[\prod_{i=2}^{n-1} \frac{V_i^{i_1...i_{n_i}}(q_i,q_{i+1})}{q_i^2} \]  \ ,
\ee
where $V_i(q_i,q_{i+1})$ are the
$RRP^{n_i}$-vertices for $n_i$-particle production in $i$-th jet.
The margin vertices defined as
\be
V_A=\frac{1}{P_A^+}\G_{||}^{\rho\n_(i)-}e_\rho(p_A)\prod e_{\n_i}(p_i^{(1)}) \ ,
V_B=\frac{1}{P_B^-}\G_{||}^{\eta\n_(j)+}e_\eta(p_B)\prod e_{\n_j}(p_j^{(n)}) \ .
\ee
For the case when a jet consists from one particle, we have omitting the color factor:
\be
V_A=g \frac{\G^{\s\eta+}(P_A,q_2)}{P_A^+} \cdot e_\s(P_A)e_\eta(P_{A'}) \ ; \ \sum (V_A)^2=8g^2 \ .
\ee
The differential cross section reads:
\be
d\s^{2 \to n} =R(s){\a_s^n 2^{-5n+4} \pi^{-2n+3}} \cdot V_A^2 V_B^2 \ \prod_{i=2}^{n-1}
\(V_i \)^2 \cdot \frac{\prod_{i=1}^{n-1} d^2{\vecc
q_i}}{\(\prod_{i=1}^{n-1}{\vecc q_i}^2\)^2} \ \cdot
\prod_{i=1}^{n} d\phi_i\ \frac{(\ln s/M^2)^{n-2}}{(n-2)!} \ ,
\ee
where the intermediate gluons reggeization factor is included:
\be
R(s)=\prod_1^{n-1}\(\frac{s_i}{M^2}\)^{2(\omega(q_i^2)-1)} \ ,
\ee
with $M$ is some averaged value of jet invariant mass and $\omega(q^2) \ , \ \omega(0)=1$
is the gluon Regge trajectory.

\section{Discussions and Conclusions}

The primary purpose of our investigation is to give a systematic self-consistent approach to derivation of the Feynman rules
directly from the effective reggeon-particle action, and, what is essential, to present the results in a most convenient form for
numerical simulations. We hope that our results will provide a practical help in writing relevant computer codes for
phenomenological needs. In this work, we present the explicit expressions for effective vertices 
of the types $PR , RP , RPP , RRP, RRPP , RPPP , RRPPP , RPPPP $ derived
directly within the framework of the effective Regge theory developed in the paper by one of us (L.N.L) \cc{L95}. Some of
these vertices were obtained by means of different methods in the Refs. \cc{LF, LPHREP, LF96, VERT}, while the two last vertices
($RRPPP$ and $RPPPP$) were not considered in literature up to now.
All presented vertices satisfy the Bose-symmetry and gauge
invariance conditions. Moreover, we made sure that the vertices of
higher orders can be constructed, step by step, using the known
ones given above.

The ``nonlinear'' vertices of type $RRRP$ with number of {\it reggeons} 
exceeding two can be constructed as well. We will not consider them here.
The program of fitting the experimental data in the QMRK using the effective
vertices given above can be realized say for extracting the information
about the gluon reggeon parameters.

\section*{Acknowledgements}

The work is partially supported by INTAS (Grant No. 00-00-366),
RFBR (Grants Nos. 04-02-16445, 03-02-17077), and Russian Federation President's Grant 1450-2003-2.
We thank V.S. Fadin, J. Bartels, R. Kirshner, B. Nicolescu, G. Salam, N.N. Nikolaev, G.P. Vacca, M.A. Braun, 
and others for helpful discussions. We are grateful to V. Bytev for help.

\section*{Appendix}

For convenience, we collect here several useful relations for
vertices, which were derived and used in this work (here it is
assumed that $k = p_1+p_2+p_3$).

Effective 4-gluon vertex entering  $RRPPP$ and $RPPPP$ has a property:
\be
p_{1\n_1}A^{\s\n_1\n_2\n_3}_{r123}=k^2 \[\frac{\tau_1}{p_{12}^2}p_1^{\n_2}g^{\s\n_3} -
\frac{\tau_2}{p_{23}^2}\g^{\n_2\n_3\s}
(-p_2,-p_3,p_{23})-\frac{\tau_3}{p_{13}^2}p_1^{\n_3}g^{\s\n_2}\] \ ,
\ee
with
$$
\tau_1=f_{12m}f_{3rm}\ , \ \tau_2=f_{1rm}f_{23m}\ , \ \tau_3=f_{13m}f_{r2m}\ , \ \sum\tau_i=0 \ .
$$

Let us demonstrate the Bose-symmetry of induced $\D_{d123e}^{\r\n_1\n_2\n_3+}(q_1,p_1,p_2,p_3,q_2)$
vertex. Really we have:
$$
D^+(q_1,d;p_1,a_1;p_2,a_2;p_3,a_3;q_2,e) = D^+(q_1,d;p_1,a_1;p_3,a_3;p_2,a_2;q_2,e) \ ; $$
\be
\frac{D^+(q_1,d;p_1,a_1;p_2,a_2;p_3,a_3;q_2,e)}{p_1^+}
=\frac{D^+(q_1,d;p_2,a_2;p_1,a_1;p_3,a_3;q_2,e)}{p_2^+} \ . \label{53}
\ee
The first relation is obvious by construction. To show the validity of the
second one, let us note that the expression
$$
\frac{1}{p_2^+}\[\frac{1}{p_{13}^+}\(\frac{\beta_1}{p_1^+}-\frac{\beta_2}{p_3^+}\)+
\frac{1}{p_{12}^+}\(\frac{\beta_3}{q_1^+}+\frac{\beta_4}{p_3^+} \) +
\frac{1}{p_{23}^+} \(\frac{\beta_5}{q_1^+}+\frac{\beta_6}{p_1^+}\) \] \ ,
$$
where
$$
\beta_1=f_{md2}f_{e1r}f_{3mr}\ ;\quad \beta_2=f_{md2}f_{3er}f_{1mr};
\quad \beta_3=f_{m21}f_{der}f_{3mr}\ ; $$
\be
\beta_4=f_{m21}f_{e3r}f_{dmr}\ ;\quad \beta_5=f_{m23}f_{der}f_{1mr};
\quad \beta_6=f_{m23}f_{e1r}f_{dmr} \ .
\ee
which is indeed the r.h.s. of the second equation, can be rearranged
using the conservation law $q_1^+=p_1^++p_2^++p_3^+$ and the relations
$$
\beta_1=-\alpha_2+\alpha_4+\alpha_5+\alpha_6 \ ; \ \beta_2=\alpha_2-\alpha_4 \ ; \
\beta_3=-\alpha_3 \ ; \ \beta_4=-\alpha_4 \ ; \ \beta_5=-\alpha_3+\alpha_5 \ ; \ \beta_6=
\alpha_1+\alpha_2+\alpha_3-\alpha_5 \
$$
giving the l.h.s. of the Eq. (\ref{53}).

For testing the gauge properties of margin vertices some identities can be used:
\be
x_1\gamma^{\n_2\n_3\rho}(-p_2,-p_3,p_{23})+x_2\gamma^{\n_3\rho\n_2}(p_2-P_A,P_A,-p_2)-
x_3\gamma^{\n_2\rho\n_3}(p_3-P_A,P_A,-p_3)+\gamma^{\rho\n_2\n_3\s}_{d23r}(p_1-q_2)_\s = 0 \ ,
\ee
with
\be
x_1=f_{32m}f_{drm}\ , \ x_2=f_{2dm}f_{3rm} \ , \ x_3=f_{d3m}f_{2rm}\ , \ x_1+x_2+x_3=0 \ ,
\ee
and, besides this:
\be
f_{1em}f_{d2r}f_{m3r} -
f_{13r}f_{2dm}f_{erm} - f_{rd1}f_{r2m}f_{e3m} +
f_{12r}f_{e3m}f_{drm} = 0 \ ;
\ee
\be
f_{m1e}\gamma^{-\n_2\n_3+}_{d23m}+f_{m1d}\gamma^{-\n_2\n_3+}_{m23e}+f_{m12}\gamma^{-\n_2\n_3+}_{dm3m}+
f_{m13}\gamma^{-\n_2\n_3+}_{d2me}=0 \ .
\ee

\newpage

\begin{figure}
\begin{center}
\includegraphics[width=.9\hsize ]{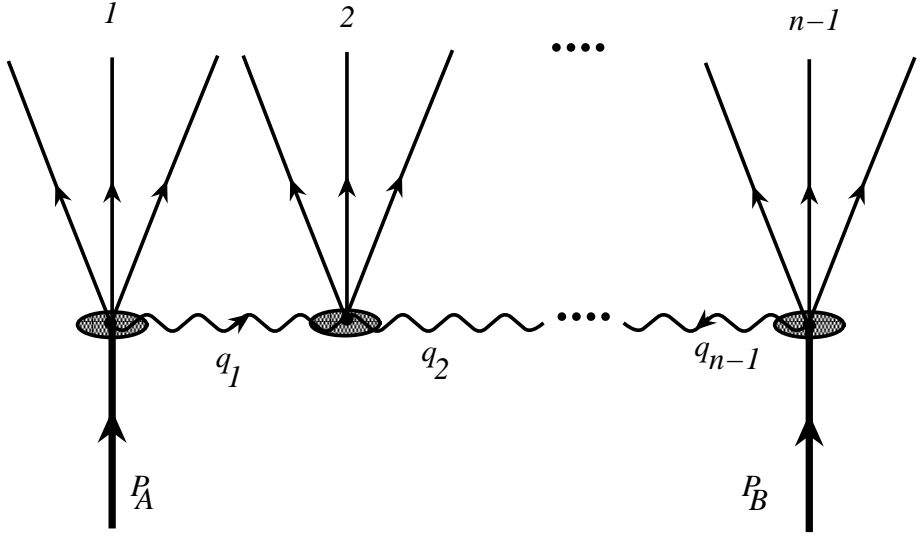}
\caption{Quasi-multi-Regge kinematics: notations. Process of
$n$-jet production $2 \to n_{jets}$ in the quasi-multi-Regge
kinematics.}
\end{center}
\end{figure}

\begin{figure}
\begin{center}
\includegraphics[width=1.0\hsize ]{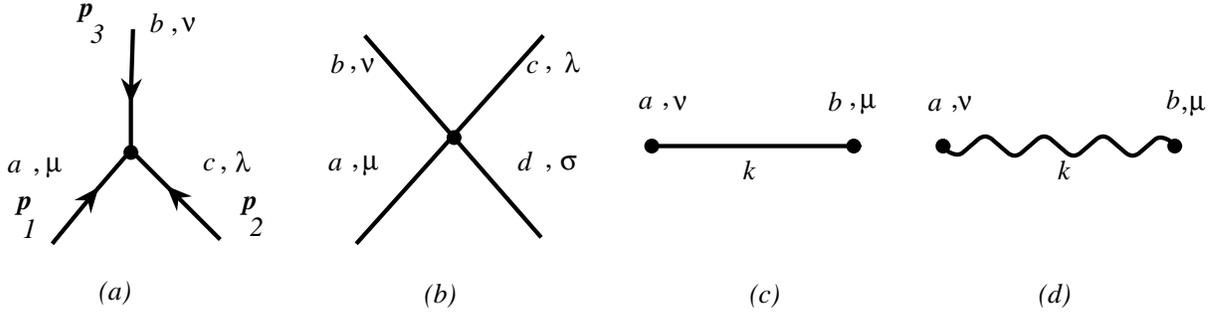}
\caption{Standard QCD Feynman rules: vertices (a-c), and the
reggeized gluon propagator (d).}
\end{center}
\end{figure}

\begin{figure}
\begin{center}
\includegraphics[width=.9\hsize ]{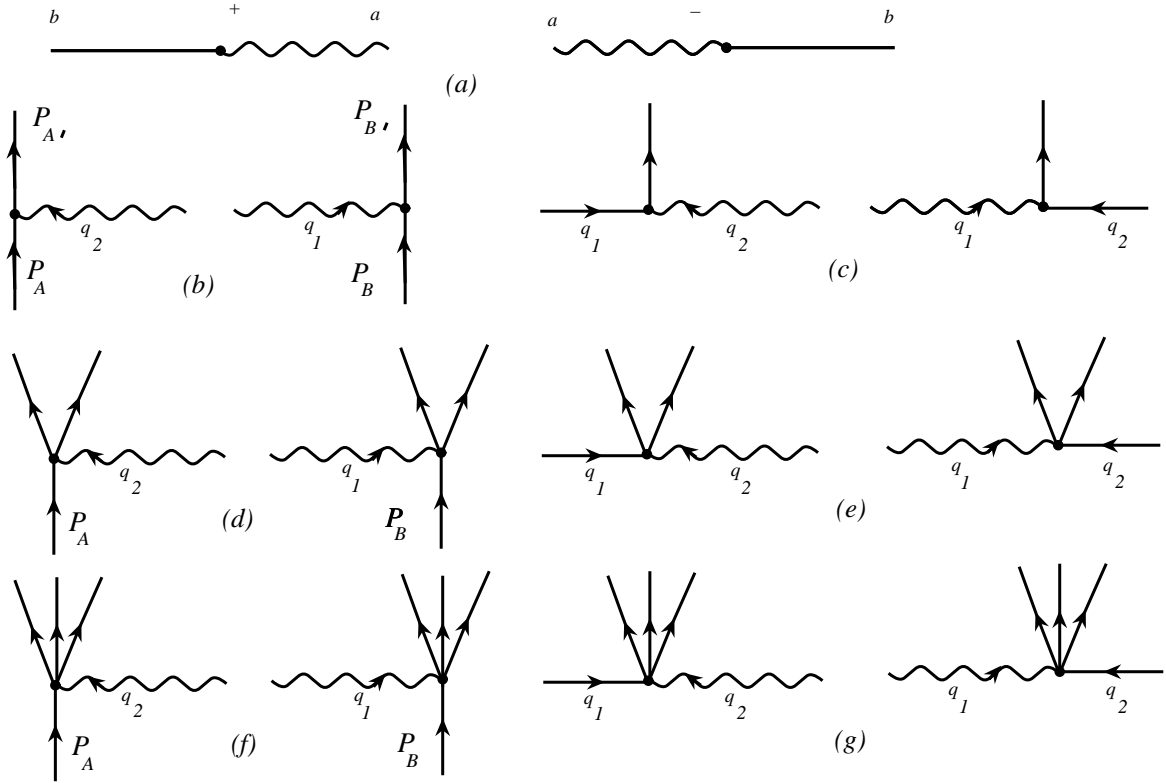}
\caption{List of the induced vertices.}
\end{center}
\end{figure}

\begin{figure}
\begin{center}
\includegraphics[width=1.\hsize ]{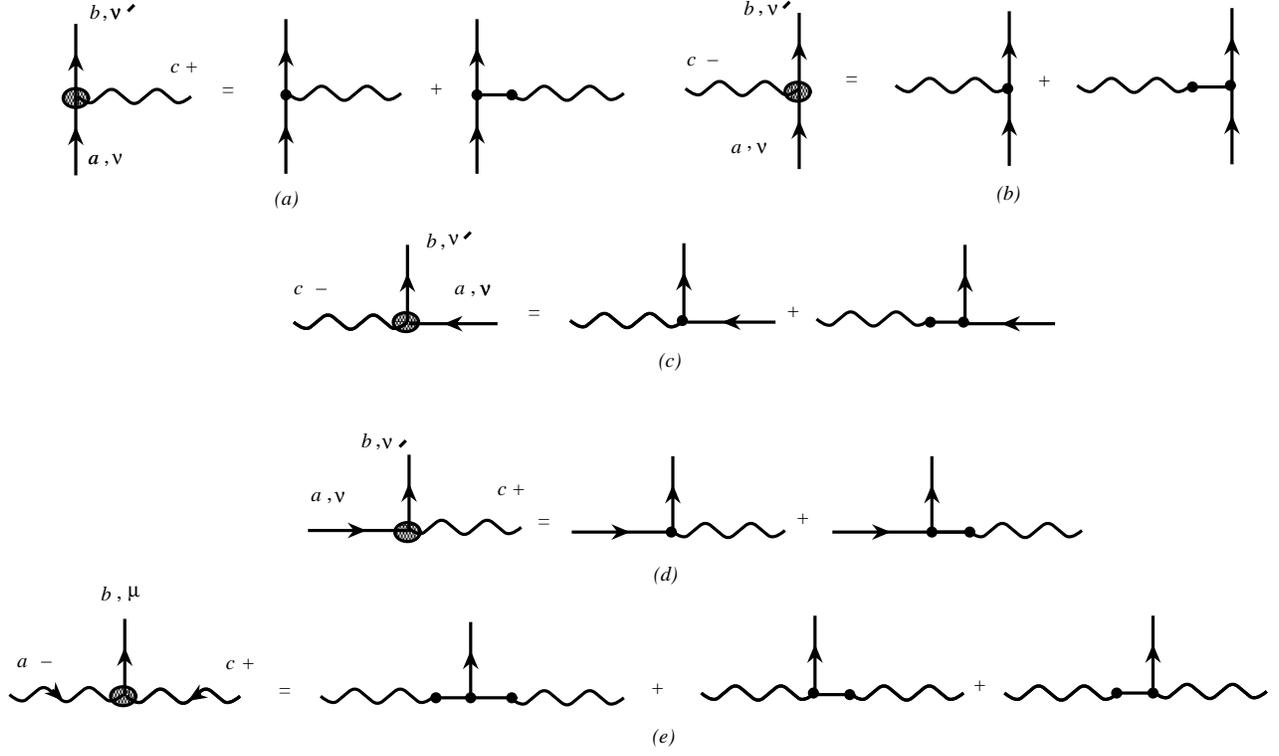}
\caption{PPR (a-d) and RRP (e) effective vertices.}
\end{center}
\end{figure}

\begin{figure}
\begin{center}
\includegraphics[width=.9\hsize ]{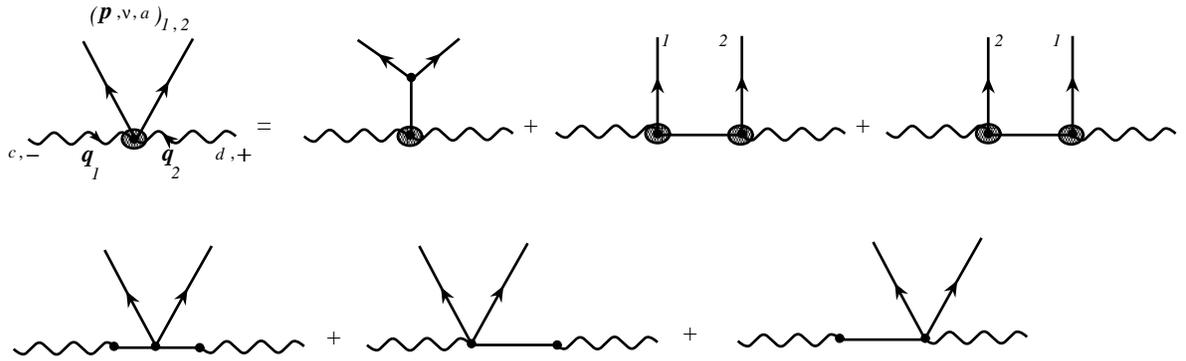}
\caption{Central $PPRR$ effective vertex.}
\end{center}
\end{figure}

\begin{figure}
\begin{center}
\includegraphics[width=.9\hsize ]{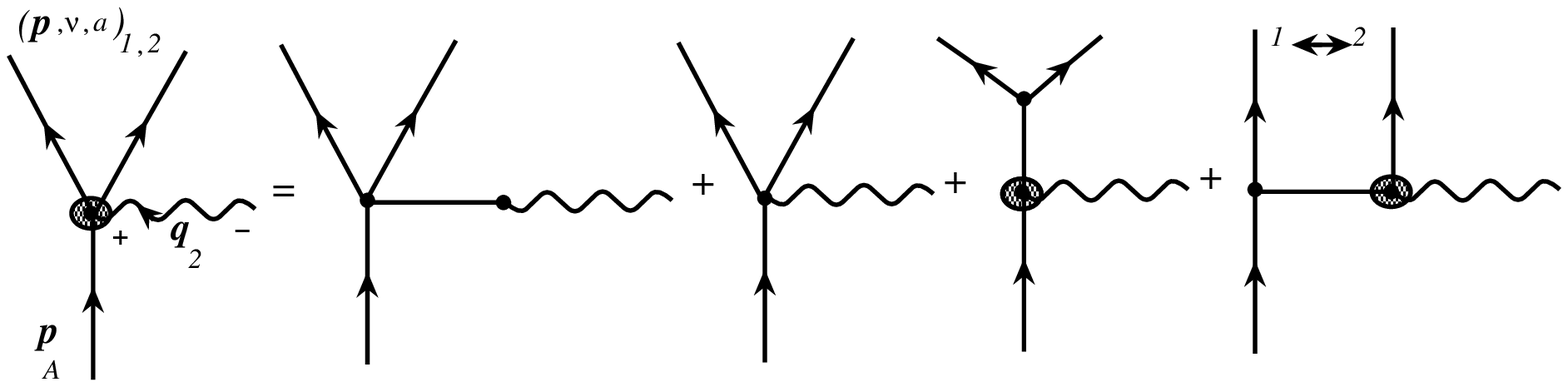}
\caption{Margin $PPPR$ effective vertex: ``left'' type.}
\end{center}
\end{figure}

\begin{figure}
\begin{center}
\includegraphics[width=.9\hsize ]{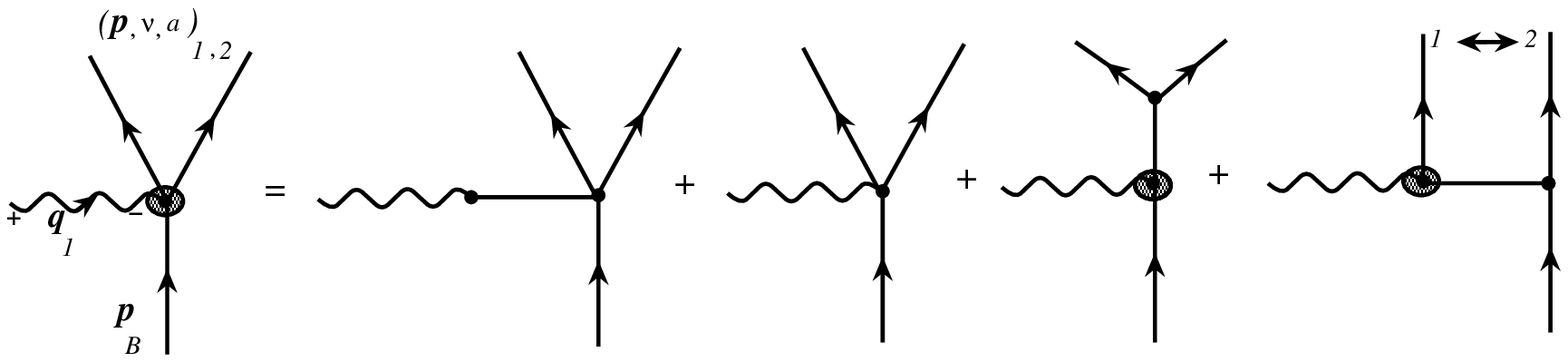}
\caption{Margin $PPPR$ effective vertex: ``right'' type.}
\end{center}
\end{figure}

\begin{figure}
\begin{center}
\includegraphics[width=1.\hsize ]{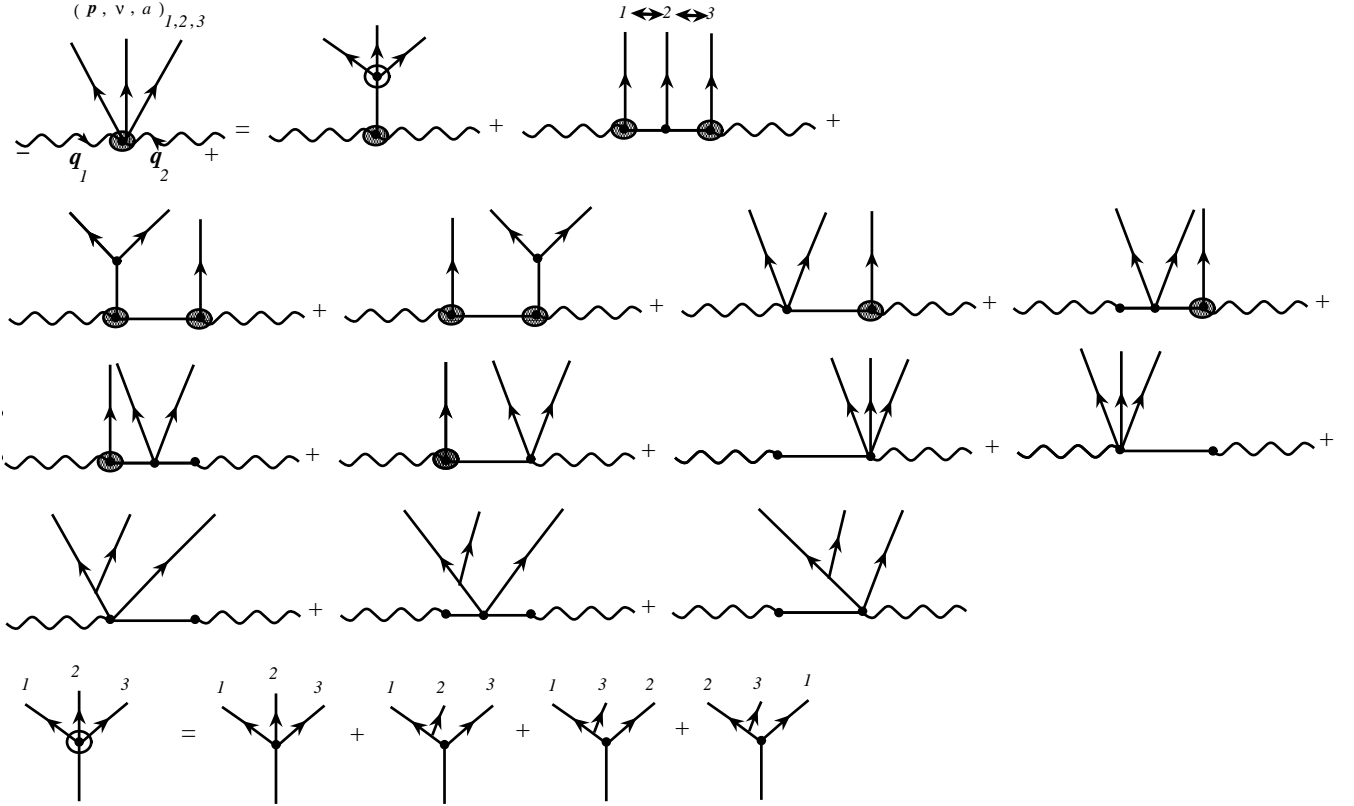}
\caption{Central $RRPPP$ effective vertex.}
\end{center}
\end{figure}

\begin{figure}
\begin{center}
\includegraphics[width=.9\hsize ]{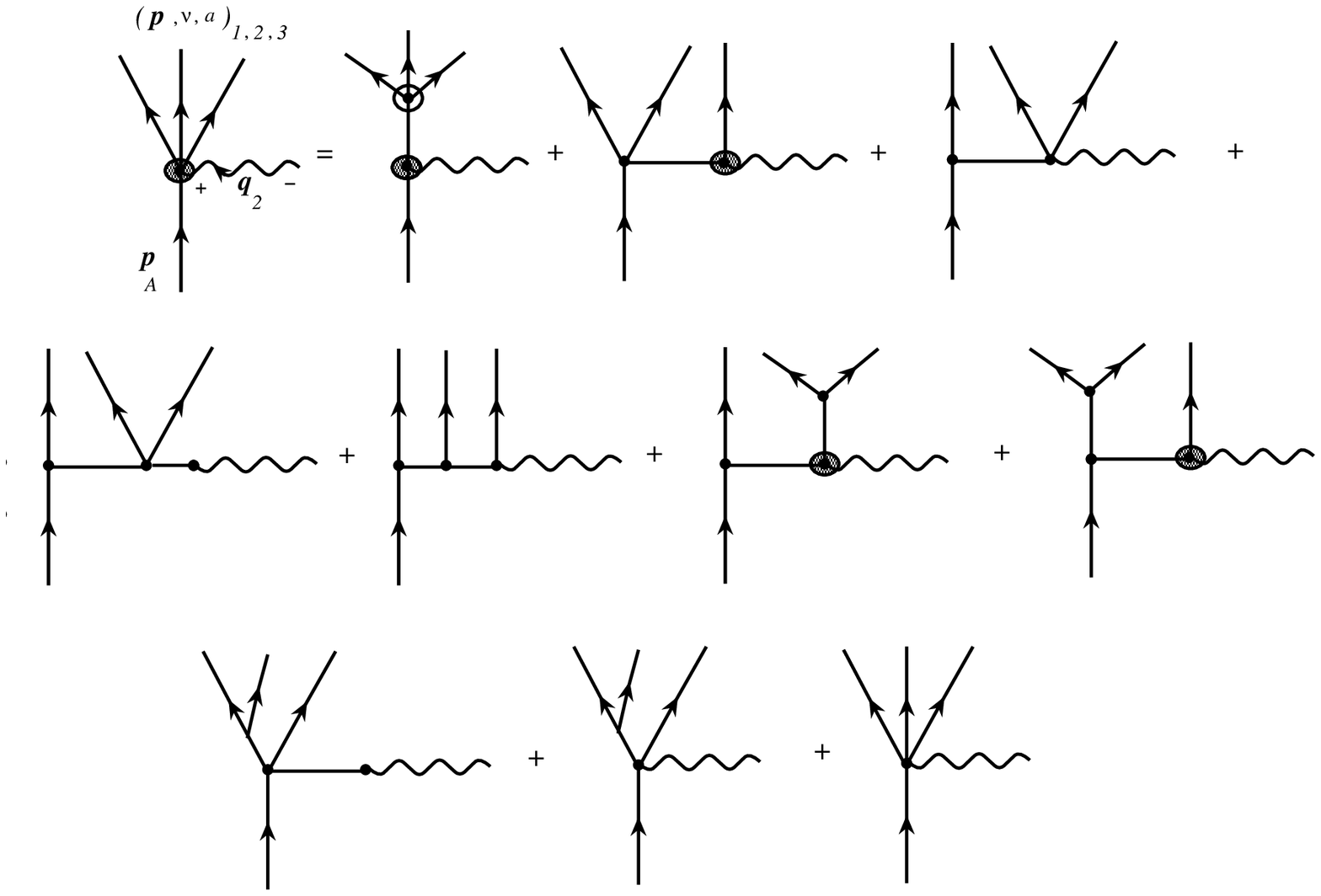}
\caption{Margin $PPPPR$ effective vertex: ``left'' type.}
\end{center}
\end{figure}

\begin{figure}
\begin{center}
\includegraphics[width=0.9\hsize]{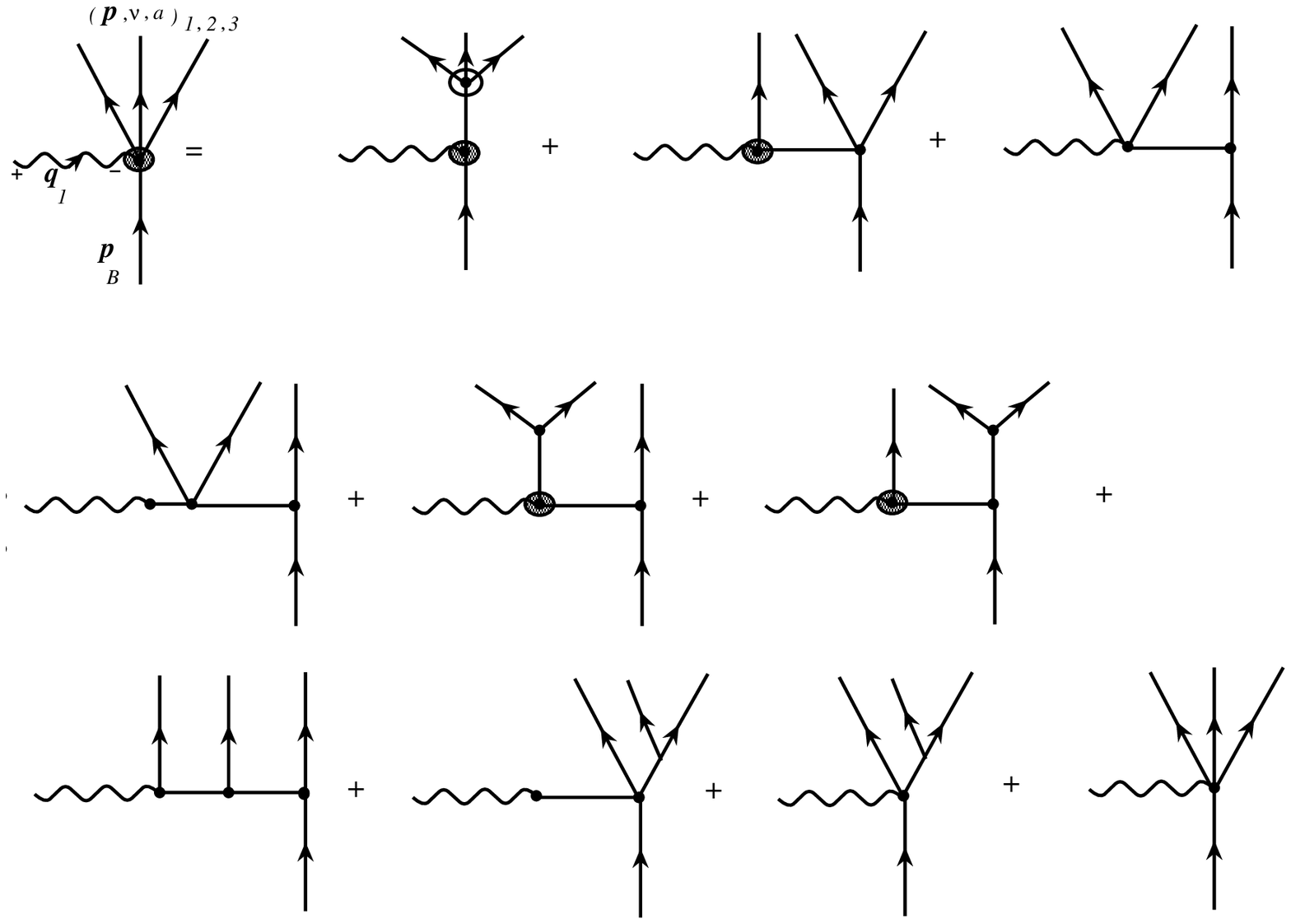}
\caption{Margin $PPPPR$ effective vertex: ``right'' type.}
\end{center}
\end{figure}

\end{document}